\documentclass[]{aastex631}

\usepackage{algpseudocode}

\begin{document}

\title{A kinetic-magnetohydrodynamic model with adaptive mesh refinement for modeling heliosphere neutral-plasma interaction}

\author{Yuxi Chen}
\affiliation{Boston University}

\author{Gabor Toth}
\affiliation{University of Michigan}

\author{Erick Powell}
\affiliation{Boston University}

\author{Talha Arshad}
\affiliation{University of Michigan}

\author{Ethan Bair}
\affiliation{Boston University}

\author{Marc Kornbleuth}
\affiliation{Boston University}

\author{Merav Opher}
\affiliation{Boston University}



\begin{abstract}
The charge exchange between the interstellar medium (ISM) and the solar wind plasma is crucial for determining the structures of the heliosphere. Since both the neutral-ion and neutral-neutral collision mean free paths are either comparable to or larger than the size of the heliosphere, the neutral phase space distribution can deviate far away from the Maxwellian distribution. A kinetic description for the neutrals is crucial for accurately modeling the heliosphere. It is computationally challenging to run three-dimensional (3D) time-dependent kinetic simulations due to the large number of macro-particles. 
In this paper, we present the new highly efficient SHIELD-2 model with a kinetic model of neutrals and a magnetohydrodynamic (MHD) model for the ions and electrons. To improve the simulation efficiency, we implement adaptive mesh refinement (AMR) and particle splitting and merging algorithms for the neutral particles to reduce the particle number that is required for an accurate simulation. We present several tests to verify and demonstrate the capabilities of the model.
 \end{abstract}



\section{Introduction} \label{sec:intro}

The interaction between the interstellar medium (ISM) and the solar wind is a fundamental process that shapes the heliosphere. The charge exchange between the ISM neutrals and solar wind protons slows down the solar wind and has a significant impact on the global structures of the heliosphere, such as the location of the termination shock \citep{baranov1993model}. The charge exchange produces energetic neutral atoms (ENAs), which can be detected at Earth's orbit and used as a remote sensing tool for inferring the global structures of the heliosphere. A lot of works have been done with the ENA data from the Interstellar Boundary Explorer (IBEX) \citep{schwadron2014separation, mccomas2024fourteen}, for example, \citet{zirnstein2020distance} investigated the distance from the Sun and the ENA source, and \citet{kornbleuth2023probing} tried to determine the length of the heliotail.

Although the solar wind proton-proton collision mean free path is large, the plasma phase space distribution can be relaxed to a Maxwellian efficiently by wave-particle interactions, so it is valid to treat the plasma as a fluid at the scale of the heliosphere. The mean free path of the neutrals is on the order of the heliosphere scale \citep{izmodenov2000hot}, and there is no collisionless mechanism to thermalize the neutrals. A kinetic description for the neutrals is therefore necessary for accurate modeling of the heliosphere Due to the limitations of computational resources, the first numerical model for ISM-solar wind interaction by \citet{baranov1981three} treated the neutrals as a single Maxwellian fluid. Since the neutrals generated by charge exchange in different regions of the heliosphere may have very different properties, later models simulate these neutrals separately with multi-fluid equations \citep{Opher:2009, zank1996interaction, alexashov_kinetic_2005}. However, each neutral fluid is still assumed to follow a Maxwellian distribution. Since the neutrals are collisionless, a kinetic description for the neutrals is required for properly modeling the heliosphere. \citet{malama_monte-carlo_1991} proposed an algorithm for simulating heliosphere neutrals on a two-dimensional (2D) axisymmetric mesh, and it was implemented into the model by \citet{baranov1993model}. Since a model with a two-dimensional (2D) grid is much more computationally efficient than a full three-dimensional (3D), the 2D axisymmetric mesh is also adopted by numerous other studies \citep{lipatov_interaction_1998,muller_self-consistent_2000,heerikhuisen_interaction_2006,alexashov_kinetic_2005}. 3D models have also been developed by \citet{pogorelov2008probing}, \citet{izmodenov_three-dimensional_2015} and \citet{michael_solar_2022}. We note that among these models, \citet{baranov1993model} is only capable of solving stationary problems, and \citet{izmodenov_solar_2005} extended its capability to support time-dependent simulations. The algorithms employed by \citet{lipatov_interaction_1998} support time-dependent simulations in principle. In practice, it is extremely challenging to run 3D time-dependent simulations since a large number of macro-particles are required to reduce statistical noise.

\citet{michael_solar_2022} introduced the Solar Wind with Hydrogen Ion Exchange and Large-scale Dynamics (SHIELD) model,  which couples the kinetic model Adaptive Mesh Particle Simulator (AMPS) \citep{tenishev2021application} with the magnetohydrodynamic (MHD) model BATS-R-US through the Space Weather Modeling Framework (SWMF). This paper is a follow-up work of \citet{michael_solar_2022}, and we introduce the updated SHIELD-2 model, which replaces AMPS with the Flexible Exascale Kinetic Simulator (FLEKS) \citep{chen_fleks_2023} as the kinetic particle tracker (PT) component. To make 3D time-dependent kinetic simulations feasible, we should control the statistical noise to an acceptable low level while keeping the total number of macro-particles as low as possible. To achieve this goal, we adopt a grid with adaptive mesh refinement (AMR) for the particle tracker to reduce the total number of cells and macro-particles that are required for resolving regions of interest. Since the charge exchange keeps generating new macro-particles, and the AMR grid would introduce uneven particle number distribution across the grid resolution change boundaries, efficient and accurate particle splitting and merging algorithms are implemented for controlling the number of particles per cell (ppc). Compared to \citet{michael_solar_2022}, we also utilize a new algorithm for accumulating charge exchange sources to reduce the statistical noise. These new features of SHIELD-2 are described in section~\ref{sec:model}. The numerical validation of the model is presented in section~\ref{sec:validation}, and we summarize it in section~\ref{sec:summary}.

\section{Model Description} \label{sec:model}
SHIELD-2 adopts a hybrid approach, modeling the plasma as a fluid and simulating neutrals kinetically using macro-particles. The following sub-section provides an overview of the model, including the governing equations for the charge exchange process and the corresponding numerical algorithms.

\subsection{Overview of the kinetic-MHD model}
\label{sec:overview}
The MHD code BATS-R-US \citep{Powell:1999} forms the backbone of our outer heliosphere (OH) model. It can run either independently, simulating both plasma and neutrals as fluids, or only simulate the plasma fluid and obtain the kinetic charge exchange sources by coupling to a kinetic particle tracking (PT) code. In the work of \citet{michael_solar_2022}, the code Adaptive Mesh Particle Simulator (AMPS) was used as the PT component. The work presented in this paper utilizes a different code, the FLexible Exascale Kinetic Simulator (FLEKS), originally designed for particle-in-cell (PIC) simulations \citep{chen_fleks_2023}. The OH component BATS-R-US and the PT component FLEKS are coupled through the interfaces provided by the Space Weather Modeling Framework (SWMF) \citep{Toth:2005swmf, Toth:2012swmf}. 

\begin{figure}
    \centering
    \includegraphics[width=0.8\textwidth]{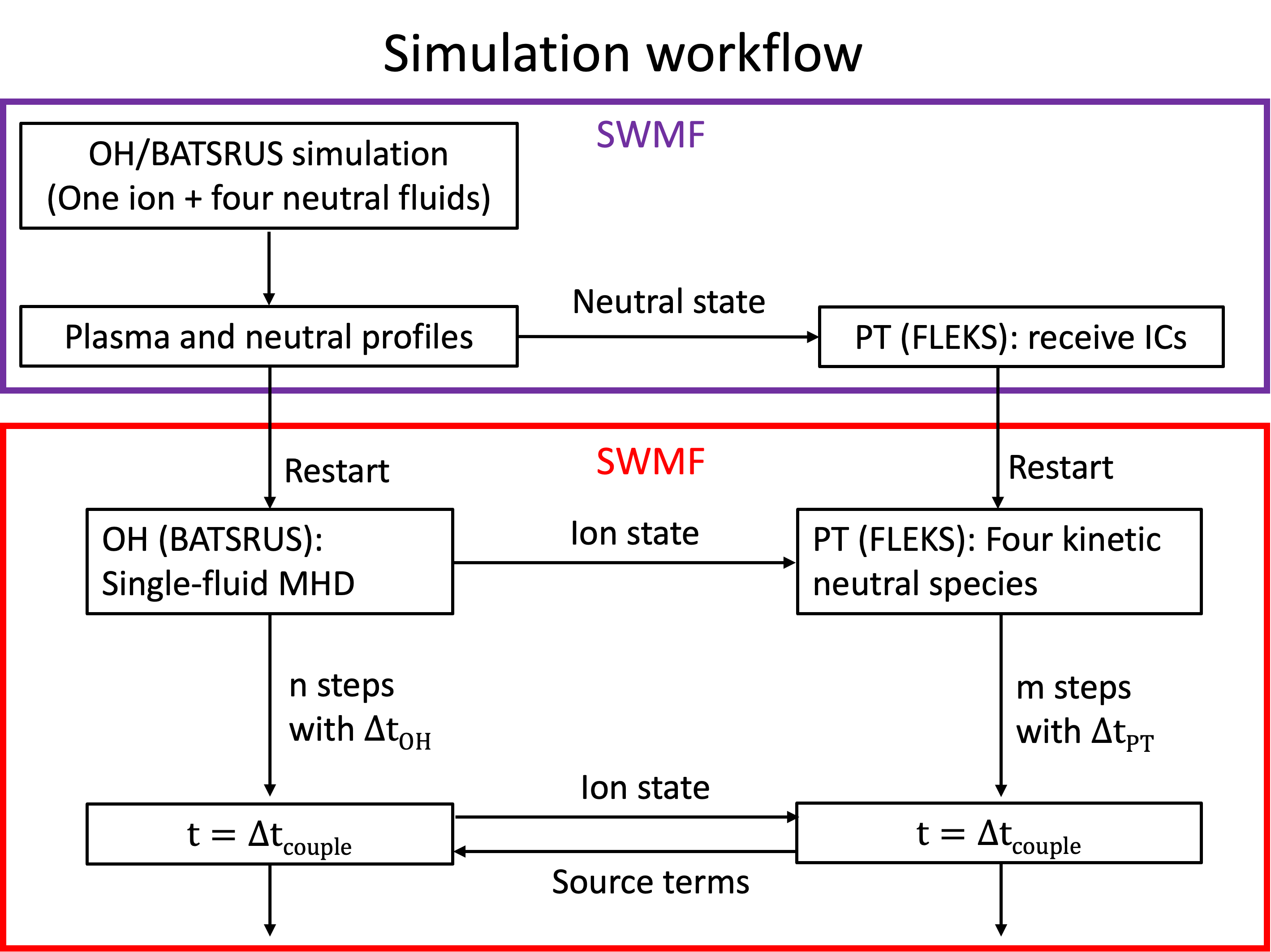}
    \caption{The workflow of a typical kinetic-MHD heliosphere simulation.}
    \label{fig:workflow}
\end{figure}

Figure~\ref{fig:workflow} illustrates the workflow of a typical kinetic-MHD heliosphere simulation. We first run the standalone BATS-R-US with five fluids, i.e., one plasma fluid and four neutral fluids, to obtain a steady-state solution. We refer readers to \citet{Opher:2009} for the details of the multi-fluid model. We then run the coupled kinetic-MHD simulation, in which BATS-R-US only simulates the plasma fluid, and the four neutral fluids are modeled by FLEKS with macro-particles. We note that the initial conditions of both BATS-R-US (OH) and FLEKS (PT) are obtained from the aforementioned steady-state solution. Neutral macro-particles are initialized with Maxwellian distributions. During initialization, the OH component also sends the plasma fluid information needed for calculating charge exchange rates to the PT component. Both components then update independently with their own time steps until the next coupling point, when updated plasma fluid information is sent from OH to PT, and charge exchange sources are sent from PT to OH. This cycle repeats until the simulation ends. We note that before the first coupling from PT to OH, the source terms for the plasma fluid are assumed to be zero. Compared to \citet{michael_solar_2022}, our model is different in two aspects:
\begin{itemize}
    \item In SHIELD-2, the initial conditions of the PT component are obtained from the steady-state solution of the OH component, instead of propagating neutral particles from upstream to reach a steady state. Our new approach is more computationally efficient.
    \item Our coupling strategy is more straightforward. Both OH and PT components are running in a time-dependent manner and there is no need to do sub-cycling for the neutral particles.
\end{itemize}

In the PT component, we assume the gravity force and solar radiation pressure on the neutral particles are negligible, and a neutral is moving along a straight line until it experiences charge exchange. Physically, a neutral can also be ionized by either photoionization or electron impact ionization, but they are less important than charge exchange and are ignored in the current implementation. We will incorporate them into our kinetic model in the future. The equations and algorithms for the charge exchange process are discussed in the next sub-section.

\subsection{Interaction between kinetic neutrals and plasma fluid}
\label{sec:interaction}
The single-fluid MHD equations with sources from charge exchange are
\begin{eqnarray}
    \frac{\partial \rho}{\partial t} + \nabla \cdot (\rho \mathbf{u}) &= S_{\rho}  \\
    \frac{\partial (\rho \mathbf{u})}{\partial t} + \nabla \cdot \left[\rho \mathbf{u} \mathbf{u} + \mathbf{I}\left(p + \frac{B^2}{2\mu_0}\right) - \frac{\mathbf{B}\mathbf{B}}{\mu_0}\right] &= \mathbf S_{\rho\mathbf{u}} \\
    \frac{\partial \mathbf{B}}{\partial t} - \nabla \times (\mathbf{u} \times \mathbf{B}) &=0 \\
    \frac{\partial e}{\partial t} + \nabla \cdot \left[\mathbf{u} \left(e + p + \frac{B^2}{2\mu_0}\right) - \frac{\mathbf{u}\cdot \mathbf{B}\mathbf{B}}{\mu_0} \right] &= S_{e}     
\end{eqnarray}
where the energy density $e$ is
\begin{equation}
    e = \frac{p}{\gamma - 1} + \frac{1}{2}\rho \mathbf{u}^2 + \frac{1}{2\mu_0}\mathbf{B}^2.
\end{equation}
$S_{\rho}$, $\mathbf S_{\rho\mathbf{u}}$, and $S_{e}$ are the source terms for mass, momentum, and energy, respectively. The mass source term $S_{\rho}$ caused by $H-H^+$ charge exchange is always zero, and the analytic momentum and energy source terms $\mathbf S_{\rho\mathbf{u}}$ and $S_{e}$ are given by
\begin{eqnarray}
    \mathbf{S}_{\rho\mathbf{u}} &= \rho_p \rho_H \int \int |\mathbf{v}_H - \mathbf{v}_p|  \cdot  \sigma_{ex}(|\mathbf{v}_H - \mathbf{v}_p|)  \cdot  f_H  \cdot  f_p  \cdot (\mathbf{v}_H - \mathbf{v}_p) d\mathbf{v}_H d\mathbf{v}_p \label{eq:sm1} \\
    S_{e} &= \rho_p \rho_H \int \int |\mathbf{v}_H - \mathbf{v}_p|  \cdot  \sigma_{ex}(|\mathbf{v}_H - \mathbf{v}_p|)  \cdot f_H  \cdot f_p  \cdot (\frac{\mathbf{v}_H^2}{2} - \frac{\mathbf{v}_p^2}{2}) d\mathbf{v}_H d\mathbf{v}_p, \label{eq:se1}
\end{eqnarray}
where the subscript $p$ and $H$ denote the plasma and neutral particles, respectively, and $f_H$ and $f_p$ are the normalized distribution functions of the neutral and plasma particles, respectively. 
The charge exchange cross section $\sigma_{ex}$ is a function of the relative velocity between the proton and the neutral $H$ atom. In SHIELD-2, we have implemented the charge exchange cross-section formulas from both \citet{maher1977atomic} and \citet{Lindsay2005}, and we use the \citet{maher1977atomic} formula in the following numerical tests. Once the source terms are obtained, the MHD equations are solved by BATS-R-US with the standard finite volume methods. BATS-R-US provides a variety of numerical algorithms for solving the MHD equations on either Cartesian or non-Cartesian AMR grids. For outer heliosphere simulations, we usually utilize an AMR Cartesian mesh with a second-order accurate scheme. We refer readers to \citet{Toth:2012swmf} for the numerical details of the MHD solver.

In our kinetic-MHD model, the neutrals are modeled kinetically with macro-particles. When calculating the sources produced by the interaction between a macro-particle and the plasma fluid, the macro-particle can be treated as a fluid element with zero temperature, i.e., the distribution function $f_H$ in eq.\ref{eq:sm1} and eq.\ref{eq:se1} degenerates to a Dirac delta function $\delta(\mathbf{v}-\mathbf{v}_H)$. The source terms at the node of a cell can be obtained by collecting the contributions from all the macro-particles in the nearby $2\times2\times2=8$ cells (3D):
\begin{eqnarray}
    \mathbf S_{\rho\mathbf{u}}(\mathbf{x}_0) &= \rho_p \sum_{i}^{n_{neu}} \xi(\mathbf{x}_{H,i}-\mathbf{x}_0) \cdot \frac{m_i}{V}\int |\mathbf{v}_{H,i} - \mathbf{v}_p|  \cdot  \sigma_{ex}(|\mathbf{v}_{H,i} - \mathbf{v}_p|) \cdot  f_p  \cdot (\mathbf{v}_{H,i} - \mathbf{v}_p)\,d\mathbf{v}_p \label{eq:sm2} \\
    S_{e}(\mathbf{x}_0) &= \rho_p \sum_{i}^{n_{neu}} \xi(\mathbf{x}_{H,i}-\mathbf{x}_0) \cdot \frac{m_i}{V}\int |\mathbf{v}_{H,i} - \mathbf{v}_p|  \cdot  \sigma_{ex}(|\mathbf{v}_{H,i} - \mathbf{v}_p|) \cdot  f_p  \cdot \frac12 (v_{H,i}^2 - v_p^2)\,d\mathbf{v}_p \label{eq:se2}
\end{eqnarray}
where $m_i$, $\mathbf{v}_{H,i}$ and $\mathbf{x}_{H,i}$ are the mass, velocity and position of the $i$-th macro-particle, respectively, $\mathbf{x}_0$ is the node location, and $V$ is the volume of a cell. 
The plasma bulk velocity $\mathbf u_p$ and thermal speed $v_{th}$ are obtained at the location of $\mathbf{x}_{H,i}$ and they are used to calculate the shifted Maxwellian distribution $f_p \sim e^{-(\mathbf{v}_p - \mathbf{u}_p)^2/v_{th}^2}$. 
We use a second-order linear interpolation function $\xi(\mathbf{x}_{H,i}-\mathbf{x}_0)$ for calculating the contributions from the macro-particle locations to the cell node. 
Instead of calculating eq.\ref{eq:sm2} and eq.\ref{eq:se2} integrals on the fly, which is time-consuming, we precalculate the integrals for a set of discrete values of relative velocities $dv = |\mathbf{v}_{H} - \mathbf{u}_p|$ and thermal velocities $v_{th}$, and save the results in a two-parameter ($dv$ and $v_{th}$) lookup table for fast retrieval. 
Note that the momentum source term (force) for a given particle $i$ is parallel to the relative velocity $(\mathbf{v}_{H,i} - \mathbf{u}_p)$ direction due to the cylindrical symmetry of the integral over the plasma velocity space as long as the plasma velocity distribution function $f_p$ is isotropic, which is true for the Maxwellian distribution.

The expressions above are the source terms for the plasma fluid, and the associated charge exchange process also removes some neutrals and generates new neutrals. We note that a macro-particle is not equivalent to a physical particle, instead, it represents a collection of physical particles that are close to each other in the phase space. During one time step of $dt$, the statistical expectation of the mass loss of a macro-particle is
\begin{equation}
    dm_i = dt \cdot m_i \int |\mathbf{v}_{H,i} - \mathbf{v}_p|  \cdot  \sigma_{ex}(|\mathbf{v}_{H,i}-\mathbf v_p|) \cdot  f_p\, d\mathbf{v}_p,\label{eq:dw}
\end{equation}
and the new mass of the $i$-th macro-particle is $m_{i,new}=m_i - dm_i$. The $H-H^+$ charge exchange process also creates new neutrals with the same amount of the total mass $dm_i$. 
The velocity and location of this secondary neutral is the same as its parent proton. From eq.\ref{eq:sm2} and eq.\ref{eq:se2}, it is clear that new secondary neutral particles generated by the $i$-th macro-particle follow the distribution of: 
\begin{eqnarray}
    \nu(\mathbf{v}_p) &\sim |\mathbf{v}_{H,i} - \mathbf{v}_p|  \cdot  \sigma_{ex}(|\mathbf{v}_{H,i} - \mathbf{v}_p|) \cdot f_p. \label{eq:nu}
\end{eqnarray}
The rejection sampling method can be applied to draw new macro-particles from the distribution above. In a cell with $N_1$ macro-particles, all of them contribute to the sources of eq.\ref{eq:sm2} and eq.\ref{eq:se2} and adjust their masses accordingly ($m_{i,new}=m_i - dm_i$). However, we only randomly choose $N_{source}$ (usually between 1 and 8) of them for generating new macro-particles to avoid a rapid increase of the macro-particle number. Among the $N_1$ particles, the probability of the $i$-th particle being chosen is proportional to $dm_i$. After all the $N_{source}$ new particles are generated from distribution eq.\ref{eq:nu} with masses $dm_i$ (eq.\ref{eq:dw}), the mass of $i$-th new particle needs to be adjusted to $dm_i^*$ to conserve the total mass, where $dm_i^*$ is
\begin{equation}
    dm_i^*=dm_i \frac{\sum_{j = 1}^{N_1} dm_j }{\sum_{k = 1}^{N_{source}} dm_k}.
    \label{eq:dm_new}
\end{equation}
If the mass $dm_i^*$ is too small, we can adaptively adjust $N_{source}$ to save computational resources. In the simulations presented in section~\ref{sec:validation}, $N_{source}$ can be reduced to as low as 1 to keep the source particle mass above $10\%$ of the average mass $m_{avg}=\frac{\sum_{i = 1}^{N_1} m_i}{N_{init}}$, where $N_{init}$ is the initial ppc number.
In some models \citep{heerikhuisen_interaction_2006,michael_solar_2022}, the plasma fluid sources are calculated from interactions between a neutral macro-particle and a single charged particle sampled from the distribution (Eq.\ref{eq:nu}). This approach can be susceptible to statistical fluctuations. In contrast, our method calculates the sources from Eq.~\ref{eq:sm2} and Eq.~\ref{eq:se2} using a lookup table, which integrates over the entire plasma distribution function. This accounts for contributions from all plasma particles and provides a more accurate approach.

To provide physically interesting information, we label neutral macro-particles based on the region they originated from: the region between the bow shock and the heliopause (population I), the heliosheath (population II), the supersonic solar wind (population III), and the pristine ISM (population IV). We refer readers to \citet{michael_solar_2022} for more detail about the definitions of these populations. From a numerical modeling point of view, the population number is just a label of a macro-particle that is used to identify where it is generated from. A macro-particle can move into other regions and interact with the plasma fluid there. When calculating the sources (eq.\ref{eq:sm2} and eq.\ref{eq:se2}), contributions from all populations are taken into account for a given cell. For example, in a cell located within region $X$, macro-particles of all populations will charge exchange with ions and generate new macro-particles. However, these new particles will all be labeled as population $X$.

As shown in Figure~\ref{fig:workflow}, our OH and PT components are allowed to run independently with their time steps. The coupling interval $dt_{couple}$ is explicitly chosen and can differ from the time steps of both components. Assume the PT component runs $n_1$ steps during one coupling interval, the sources ($\mathbf S_{\rho \mathbf u}^*$ and $S_{e}^*$) sources that are passed from PT to OH are calculated with the following algorithm:
\begin{algorithmic}
   \State
   \State $\mathbf S_{\rho \mathbf u}^* = 0$
   \State $S_{e}^* = 0$
   \State $t^* = t$
   \While{$t^* < t + dt_{couple}$}
         \State calculate $dt$, $\mathbf S_{\rho \mathbf u}$, $S_e$
         \State $\mathbf S_{\rho u}^* = \mathbf S_{\rho \mathbf u}^* + \mathbf S_{\rho \mathbf u}*dt$
        \State $S_{e}^* = S_{e}^* + S_{e}*dt$
        \State $t^* = t^* + dt$
   \EndWhile
   \State $\mathbf S_{\rho \mathbf u}^* = \mathbf S_{\rho \mathbf u}^*/(t^*-t)$
   \State $S_{e}^* = S_{e}^*/(t^*-t)$
   \State $t = t^*$
   \State
\end{algorithmic}
where $\mathbf S_{\rho \mathbf u}$ and $S_e$ are obtained from eq.\ref{eq:sm2} and eq.\ref{eq:se2}, and $dt$ is the time step of the PT component. The meshes for PT and OH are different, and the interpolation between these two meshes is done with a second-order linear interpolation algorithm. The source terms are first calculated on the nodes of the PT grid. Then, they are interpolated to the cell centers of the OH mesh. OH uses these source terms until the next coupling time. During the coupling stage, PT also receives the updated plasma fluid information from OH. 

\subsection{Adaptive mesh refinement}
\label{sec:amr}

Although a macro-particle itself is mesh-free, it interacts with the plasma fluid solved on a mesh. Increasing the resolution of the PT mesh, which stores the plasma properties and accumulates the charge exchange sources, contributes to improved accuracy in both the plasma and neutral solutions. However, maintaining a reasonable number of particles per cell (ppc) is crucial to control statistical noise. This requires the total number of macro-particles to be roughly proportional to the total number of cells. Adaptive Mesh Refinement (AMR), which only refines regions of interest, can significantly reduce the computational cost by reducing the total number of cells and macro-particles, leading to faster simulations.

FLEKS leverages the AMReX library \citep{Zhang2019,Zhang2021}, which provides efficient parallel data structures for multi-level Cartesian grids. We implemented interfaces to call AMReX functions for creating and refining the mesh. The user can define regions of interest for refinement using various geometric shapes, such as boxes, spheres, shells, and paraboloids. Examples of AMR meshes are provided in section~\ref{sec:validation}. A refinement ratio of two is used in our implementation.

Unlike a particle-in-cell code, where solving the electromagnetic fields on the mesh introduces challenges in controlling errors near refinement boundaries, the mesh in the neutral model presented here is only used for storing plasma properties and accumulating source terms, and there are no numerical difficulties for supporting AMR. 

\subsection{Particle splitting and merging}
\label{sec:resampling}

The number of particles per cell (ppc) is a crucial parameter impacting both statistical noise and simulation performance. As the charge exchange process continuously generates new macro-particles, the number of ppc increases linearly. This can lead to a significant slowdown and eventual memory exhaustion if no mechanism is implemented to control the macro-particle number.

Adaptive Mesh Refinement (AMR) introduces additional potential issues related to ppc control. When neutral particles move between grids with different resolutions (coarse to fine or vice versa), the ppc value experiences sudden changes (reduction or increase) on the receiving grid. To address these challenges and maintain a stable ppc level, particle splitting and merging algorithms are necessary.

\subsubsection{Particle merging}
\label{sec:merging}

\citet{chen_fleks_2023} introduced a particle merging algorithm for the particle-in-cell (PIC) module of FLEKS, where $N (N>5)$ particles are merged into five new particles by adjusting the particle masses while conserving the total mass, momentum, and energy. The algorithm does not alter particle velocities so that the velocity space distribution is conserved as much as possible. A merging may fail due to the unphysical solutions (negative mass) of the linear system for the new particle masses. 
We found the larger the $N$ is, the more likely the merging fails, so the merging efficiency is limited. The algorithm works well for the PIC code on a uniform grid, where the ppc number does not change drastically during a simulation. We have tested this algorithm for our kinetic-MHD model and found it is not efficient enough to control the rapid ppc number increase due to the aforementioned reasons. 

To increase the merging success rate, we improve the algorithm by merging $N (N>M)$ particles into $M (M\ge 5)$ new particles. The new algorithm still follows the same ideal of \citet{chen_fleks_2023}, i.e., binning particles in the velocity space and merging $N$ particles, which are in the same velocity bin, into $M$ new particles by adjusting the particle masses while conserving the total mass, momentum, and energy. The conservation requirements introduce five equations, while the number of the unknowns, the masses of the $M$ new particles, can be more than five, so the linear system is underdetermined and the solution is not unique. We apply a generalized Lagrange multiplier to select the optimal solution. 

As the first step, we divide the velocity space ranging from  $(v_x-2v_{th}, v_y-2v_{th}, v_z-2v_{th})$ to $(v_x+2v_{th}, v_y+2v_{th}, v_z+2v_{th})$ into $n_{bin}=\left\lceil 0.5 N_p^{1/3}\right\rceil$ bins in each direction for each cell, where $N_p$ is the ppc number and $v_{th}$ is the thermal speed. We then select $N$ particles from each velocity bin for merging. Because new light macro-particles are generated at every step, selecting $N$ lightest particles for merging has a smaller impact on the overall velocity distribution than merging heavy particles.

The total mass, momentum, and energy of the original $N$ particles are:
\begin{equation}
    m_t = \sum_{i=1}^{N} m_i, \qquad
\mathbf{p}_t = \sum_{i=1}^{N} m_i \mathbf{v}_i, \qquad
e_t = \sum_{i=1}^{N} \frac{1}{2}m_i v_i^2,
\end{equation}
where $m_i$ and $\mathbf{v}_i$ are the mass and velocity of the $i$-th particle, respectively. Among these $N$ particles, we randomly remove $N-M$ of them, and adjust the masses of the rest $M$ particles to satisfy the conservation requirements:
\begin{eqnarray}
    m_t = \sum_{i=1}^{M} m_{i}^*, \quad
    \mathbf{p}_t &=& \sum_{i=1}^{M} m_{i}^* \mathbf{v}_i, \quad
    e_t = \sum_{i=1}^{M} \frac{1}{2}m_{i}^* v_i^2
    \label{eq:linear} \\
    m_{i}^* &>& 0, i = 1, 2,...,M \label{eq:positive}
\end{eqnarray}

Since the linear system of eq.\ref{eq:linear} is underdetermined, we choose the solution that minimizes the relative change of the masses:
\begin{equation}
f(m_1^*,...,m_M^*) = 
\sum_{i=1}^{M} \left(\frac{m_i^*}{m_i} - 1\right)^2.
\end{equation}
The corresponding Lagrange function is
\begin{equation}
    L(m_1^*,...,m_M^*, \lambda_1,...,\lambda_5) = f(m_1^*,...,m_M^*) + \sum_{i=1}^{5} \lambda_i g_i,
\end{equation}
where,
\begin{eqnarray}
   g_1(m_i^*) &=& m_t - \sum_{i=1}^{M} m_{i}^* \\
   g_2(m_i^*) &=& p_{t,x} - \sum_{i=1}^{M} m_{i}^* v_{x,i} \\
   g_3(m_i^*) &=& p_{t,y} - \sum_{i=1}^{M} m_{i}^* v_{y,i} \\
   g_4(m_i^*) &=& p_{t,z} - \sum_{i=1}^{M} m_{i}^* v_{z,i} \\
   g_5(m_i^*) &=& e_t - \sum_{i=1}^{M} \frac{1}{2}m_{i}^* v_i^2.
\end{eqnarray}

The optimal solution is obtained by solving the following linear system:
\begin{eqnarray}
    \frac{\partial L}{\partial m_i^*} &=& 0, \quad i = 1,2,...,M \\
    \frac{\partial L}{\partial \lambda_i} &=& 0, \quad i = 1,2,3,4,5.
\end{eqnarray}
Since $M$ is usually between 6 and 12, and the linear system is not large, we solve it with the direct Gaussian elimination method including pivoting. Only the physical solutions, i.e., the solutions that satisfy eq.\ref{eq:positive}, are admitted, otherwise, we skip the merging for this group of particles. To further limit the errors introduced by the merging, we add the following two extra constraints to the algorithm:
\begin{itemize}
    \item For the $N$ particles that are selected for merging, we require that the ratio between the heaviest and lightest one should not be too large, i.e., $m_{max}/m_{min} < \alpha$, where we usually choose $\alpha=10$ or $\alpha=20$. 
    \item We require the new particle mass $m_i^*$ should not be too different from the original mass $m_i$, i.e.,    
    \begin{eqnarray}
        1/\beta< \frac{m_i^*}{m_i} < \beta, \quad i = 1,2,...,M, \label{eq:beta}
     \end{eqnarray}
     where $\beta$ is a constant close to but larger than 1. In the following numerical tests, we set $\beta = 1.2$.     
\end{itemize}
If the solution does not satisfy the aforementioned constraints eq.\ref{eq:positive} and eq.\ref{eq:beta}, we can select another set of $M$ particles and try merging again. In the numerical tests shown in section~\ref{sec:validation}, we try merging at most 6 times for each group of $N$ particles.

\subsubsection{Particle splitting}
\label{sec:splitting}
A particle splitting algorithm is required for two reasons. First, the number of ppc reduces by about a factor of eight when the neutrals flow from the coarse grid side to the fine grid side, because the particle number per volume does not change too much but the volume per cell reduces by a factor of eight. Although charge exchange continuously generates new macro-particles, these new particles are typically much lighter than the original ones.  While they contribute to maintaining the overall particle population, their limited mass restricts their effectiveness in reducing statistical noise. Second, within a single cell, the masses of macro-particles can vary significantly due to the presence of large density gradients. If a few macro-particles are significantly heavier compared to others in a cell, these heavy particles dominate this cell and lead to a large statistical noise. Therefore, to address these challenges and control the statistical noise, we employ a particle splitting algorithm to specifically target and split these heavy macro-particles.

FLEKS already includes a splitting algorithm designed for PIC simulations \citep{chen_fleks_2023}. This algorithm splits one particle into two by displacing their locations a small random amount along the velocity direction while maintaining the original velocity. Due to the interaction between PIC particles and the electromagnetic fields, these two new particles will quickly diverge and follow different trajectories. However, in our neutral model, the macro-particles move along unperturbed ballistic trajectories without such interactions. 
If we directly apply the existing algorithm, the two new particles created by splitting will remain close together, offering minimal reduction in statistical noise. 
To address this limitation, we present a new particle splitting algorithm specifically tailored for the neutral model. 
Instead of displacing locations, this approach adjusts the velocities of the newly created particles. Splitting one particle into two with velocity adjustments cannot simultaneously conserve both momentum and energy. However, by splitting two particles into four, we can satisfy all conservation requirements.

The particle splitting algorithm begins by identifying candidate particles for splitting. There are two categories of particles eligible for selection:
\begin{itemize}
    \item If the number of ppc $n_{ppc}$ is lower than a threshold $n_{low}$, the heaviest $n_{low} - n_{ppc}$ particles will be chosen for splitting.
    \item Within a cell, particles with masses exceeding a threshold $m_{high}$ are targeted for splitting. Typically, $m_{high}$ is set to four times the average particle mass in the cell. This approach addresses the issue of statistical noise arising from particles with significantly larger masses compared to others.
\end{itemize}

The second step involves paring candidate particles with similar velocities for splitting. This is achieved by sorting the candidate particles in the velocity space. We first construct a velocity space grid of $N \times N \times N$ cells that can cover all the candidate particles and then assign these particles to the velocity space bins. The Morton space-filling curve (Z-order) is applied to sort the velocity bins and the particles therein. After sorting, we split every two particles into four with the following algorithm. Numerical experiments suggest that using a value of $N=8$ yields good performance.

\begin{figure}
    \centering
    \includegraphics[width=0.6\textwidth]{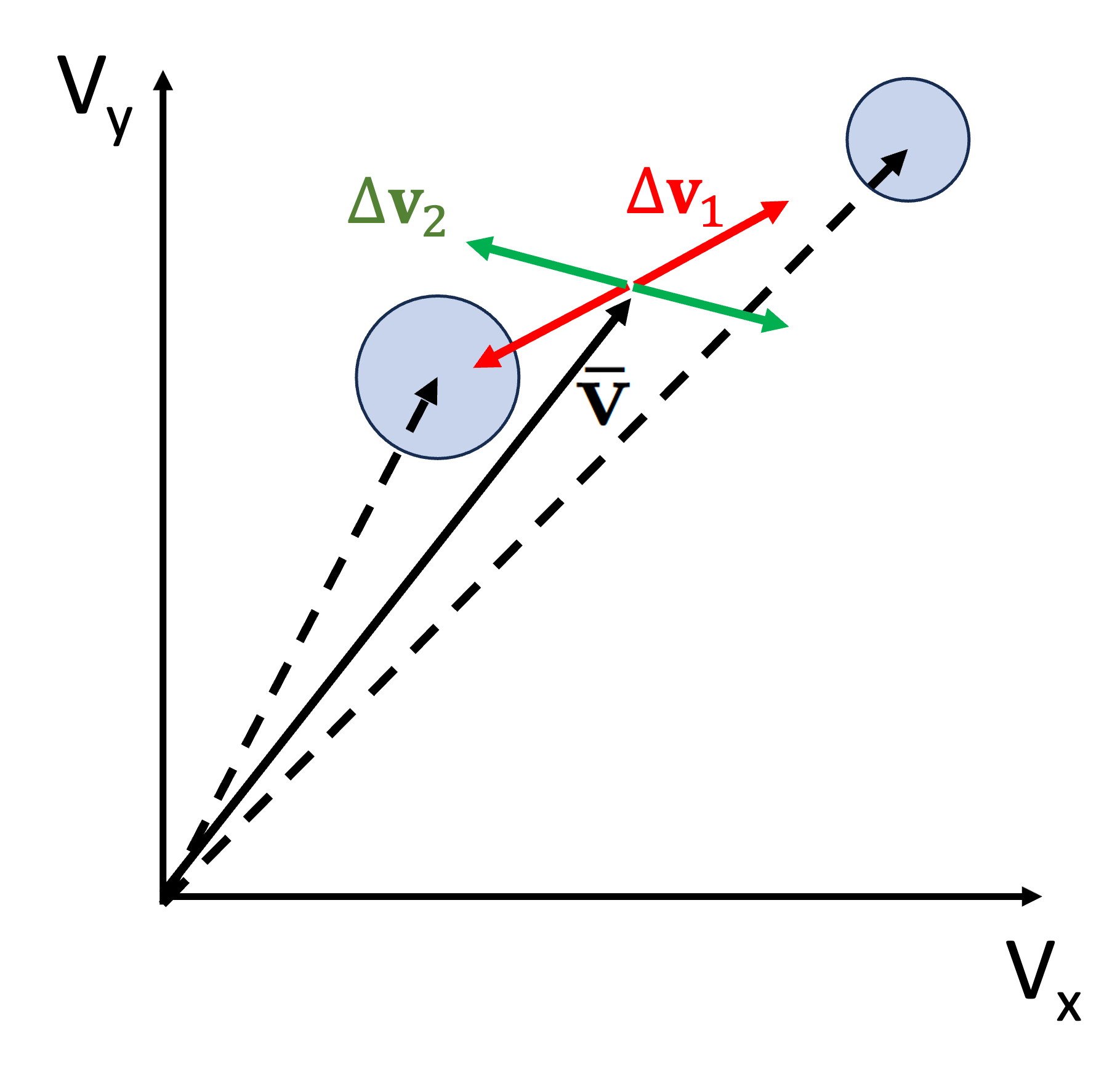}
    \caption{A diagram that illustrates the particle splitting algorithm. The two light blue circles are the original particles, and the left one is heavier. The four new particles with equal masses are created at the ends of the red and green arrows. See subsection~\ref{sec:splitting} for more detail.}
    \label{fig:split}
\end{figure}

The total mass, momentum, and energy of the original two particles are $m_t$, $\mathbf{P}_t$ and $e_t$, respectively. The average velocity is $\mathbf{\bar{v}} = \mathbf{P}_t/m_t$. We use the same weigh, i.e., $m_{new} = m_t/4$, for the new particles to satisfy mass conservation, and set their velocities as: 
\begin{eqnarray}
    \mathbf{v}_1 &=& \mathbf{\bar{v}} + \Delta \mathbf{v}_1 \\
    \mathbf{v}_2 &=& \mathbf{\bar{v}} - \Delta \mathbf{v}_1 \\
    \mathbf{v}_3 &=& \mathbf{\bar{v}} + \Delta \mathbf{v}_2 \\
    \mathbf{v}_4 &=& \mathbf{\bar{v}} - \Delta \mathbf{v}_2 
\end{eqnarray}
The momentum is conserved for any $\Delta \mathbf{v}_1$ and $\Delta \mathbf{v}_2$. Energy conservation requires that the kinetic energy due to the velocity differences relative to the average velocity does not change: 
\begin{eqnarray}
  m_{new} (\Delta \mathbf{v}_1^2 +  \Delta \mathbf{v}_2^2) = e_t - \frac12 m_t \mathbf{\bar{v}}^2.
\end{eqnarray}
The solution for $\Delta \mathbf{v}_1$ and $\Delta \mathbf{v}_2$ is not unique, and we choose a solution where the amplitudes $\Delta v_1=\Delta v_2$ are equal. The direction of $\Delta \mathbf v_1$ is chosen to be parallel with the velocity difference $\Delta \mathbf v_{orig}$ of the original two particles, while the direction of $\Delta\mathbf v_2$ is random with an isotropic distribution: 
\begin{eqnarray}
   \Delta v &=& \frac{2 e_t}{m_t} - {\bar v}^2, \\
   \Delta \mathbf{v}_1 &=& \Delta v\frac{\Delta \mathbf v_{orig}}{|\Delta v_{orig}|},  \\
   \Delta \mathbf{v}_2 &=& \Delta v\,\mathbf{e}_{random},
\end{eqnarray} 
where $\mathbf{e}_{random}$ is a unit vector with random orientation. With the solution above, the amplitude of the perturbation $\Delta v$ is proportional to the velocity difference between the two original particles. A random direction $\mathbf{e}_{random}$ is chosen for $\Delta \mathbf{v}_2$ to avoid any bias. Figure~\ref{fig:split} illustrates the splitting algorithm.

\section{Simulation results} \label{sec:validation}
\subsection{Interaction between uniform plasma and neutral flows}
\label{sec:uniform}

\begin{figure}
    \centering
    \includegraphics[width=1.0\textwidth]{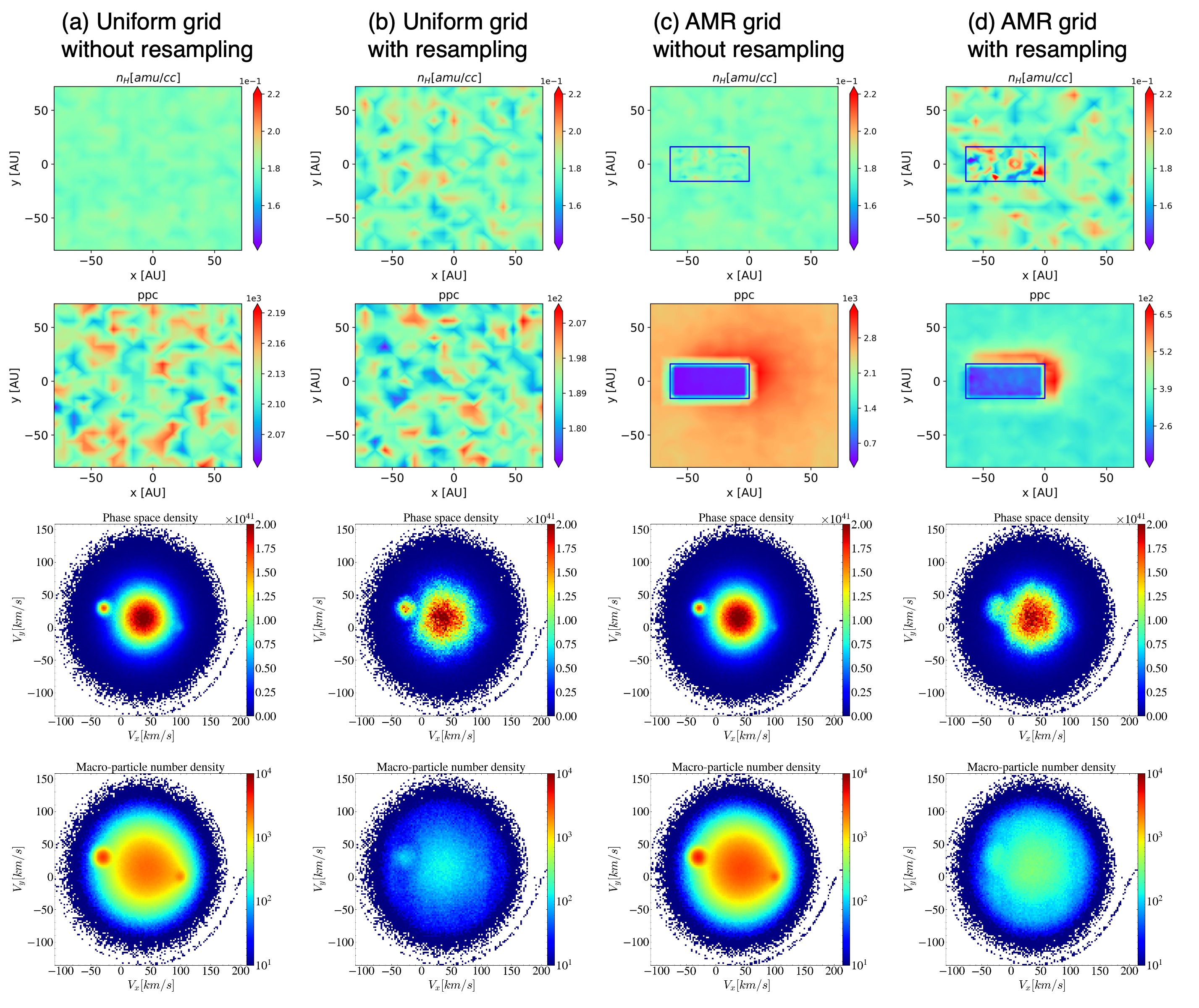}
    \caption{Results from four simulations of uniform plasma-neutral flow interaction. From top to bottom, the neutral number density, the ppc number, the phase space ($v_x - v_y$ space) density distribution, and the phase space macro-particle number density distribution are shown. Columns (c) and (d) show the simulations with an AMR mesh. The blue boxes in the first two rows enclose the area of refinement. We note the color bars for the ppc number (the second row) are different.}
    \label{fig:uniform_flow}
\end{figure}

We use a test with uniform plasma and neutral flows to validate our numerical implementation. The initial conditions of the uniform plasma flow are: $n= 0.18\,$ amu/cc, $T = 6441\,$K, and $\mathbf{u} = (100, 0, 0)\,$km/s. The neutral flow parameters are: $n = 0.18\,$amu/cc, $T = 6519\,$K, and $\mathbf{u} = (-30, 30, 0)\,$km/s. The simulation domains for both PT and OH are the same spanning from $(-80, -80, -8)\,$ AU to $(80, 80, 8)\,$AU. The grid resolution of OH is $\Delta x = 4\,$AU. The base grid resolution of PT is $\Delta x = 8\,$AU, and it can be refined to $\Delta x = 4\,$AU in the region from $(-64, -16, -4)\,$AU to $(0, 16, 4)\,$AU (blue box in 
Figure~\ref{fig:uniform_flow}). Periodic boundary conditions are applied for both components in all three directions. The MHD equations (OH component) are solved with a standard second-order finite volume method. We run the simulations for 20 years with a fixed PT time step of $\Delta t_{PT}= 0.01\,$year. 
OH and PT are coupled every $0.1~\text{year}$. Initially, 125 macro-particles per cell are launched to represent the neutrals. At every step, $N_{source} = 1$ new source macro-particle is generated for each cell. 

To validate the particle resampling algorithms and AMR implementation, we conducted four simulations using various combinations of resampling algorithms and AMR: with neither, with resampling only, with AMR only, and with both. When active,
the particle merging (splitting) algorithm is activated in cells with a ppc number that is larger (lower) than $n_{high}=188=\lceil 125 \times 1.5\rceil$ ($n_{low}=100=125\times 0.8$) threshold. 
For the merging, we combine 10 macro-particles into 8. 

The simulation results at $t=20$\,year are shown in Figure~\ref{fig:uniform_flow}. From top to bottom, the neutral number density, the ppc number, the $v_x - v_y$ phase space  density distribution, and the phase space macro-particle number density distribution are shown. 
The color bars for each row are the same except for the second row (ppc number plots). Without the particle resampling algorithms, there will be 2125 ppc at $t=20~\text{years}$ (2000 steps) on average, as shown in the second row of Figure~\ref{fig:uniform_flow}(a). The particle merging algorithm successfully maintained the ppc number at around 190 (Figure~ \ref{fig:uniform_flow}(b)). Although the noise level in Figure~\ref{fig:uniform_flow}(b) is higher than Figure~\ref{fig:uniform_flow}(a) due to the low ppc number and the errors introduced by merging, the pattern of the phase space distribution is still preserved (third row of Figure~\ref{fig:uniform_flow}). The AMR mesh causes drastic changes in the ppc number. As shown in Figure~\ref{fig:uniform_flow}(c), the ppc number reaches about 3000 on the coarse grid and the number is about 600 on the fine grid in the simulation without resampling (column (c)). In the simulation with resampling (column (d)), the ppc number is reduced by about one order of magnitude, although the merging is not efficient enough to reduce the ppc number to $n_{high} = 188$. The phase space distribution is also well preserved.

Initially, the neutral particles are assumed to be Maxwellian and the analytic momentum and energy exchange rates for the first time step can be calculated from eq.\ref{eq:sm1} and eq.\ref{eq:se1}. We compare and confirm that the first step numerical results converge to the analytic results. 

\subsection{Outer heliosphere neutral-plasma interaction}
\label{sec:heliosphere}

\begin{figure}
    \centering
    \includegraphics[width=0.8\textwidth]{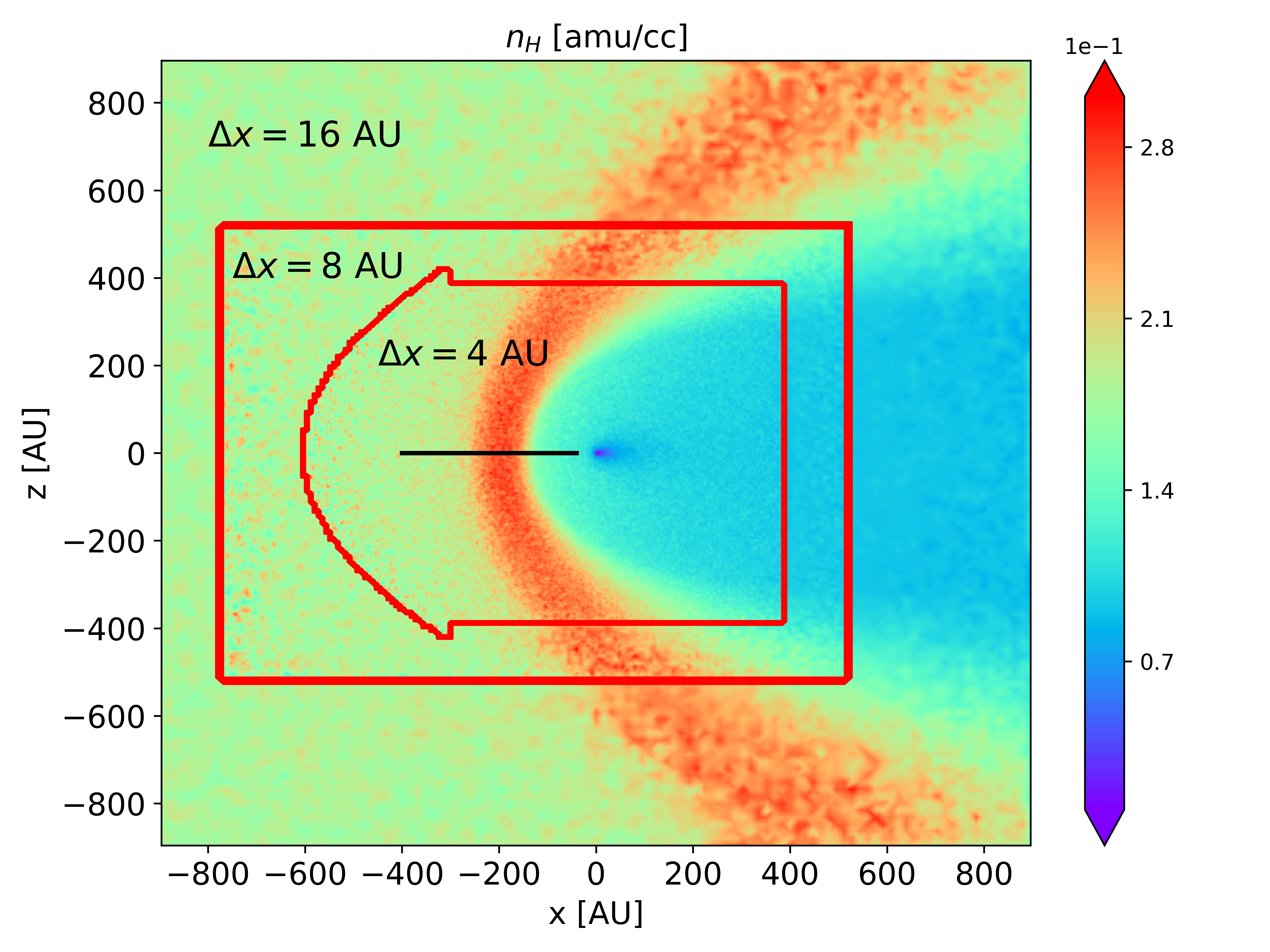}
    \caption{The total neutral density in the $y=0$ plane. The base grid with $\Delta x = 16\,$AU is refined twice to reach the finest grid resolution of $\Delta x = 4\,$AU. The red lines indicate the boundaries of AMR resolution change. The simulation results along the black line are shown in Figure~\ref{fig:lines}.}
    \label{fig:amr}
\end{figure}

\begin{figure}
    \centering
    \includegraphics[width=0.8\textwidth]{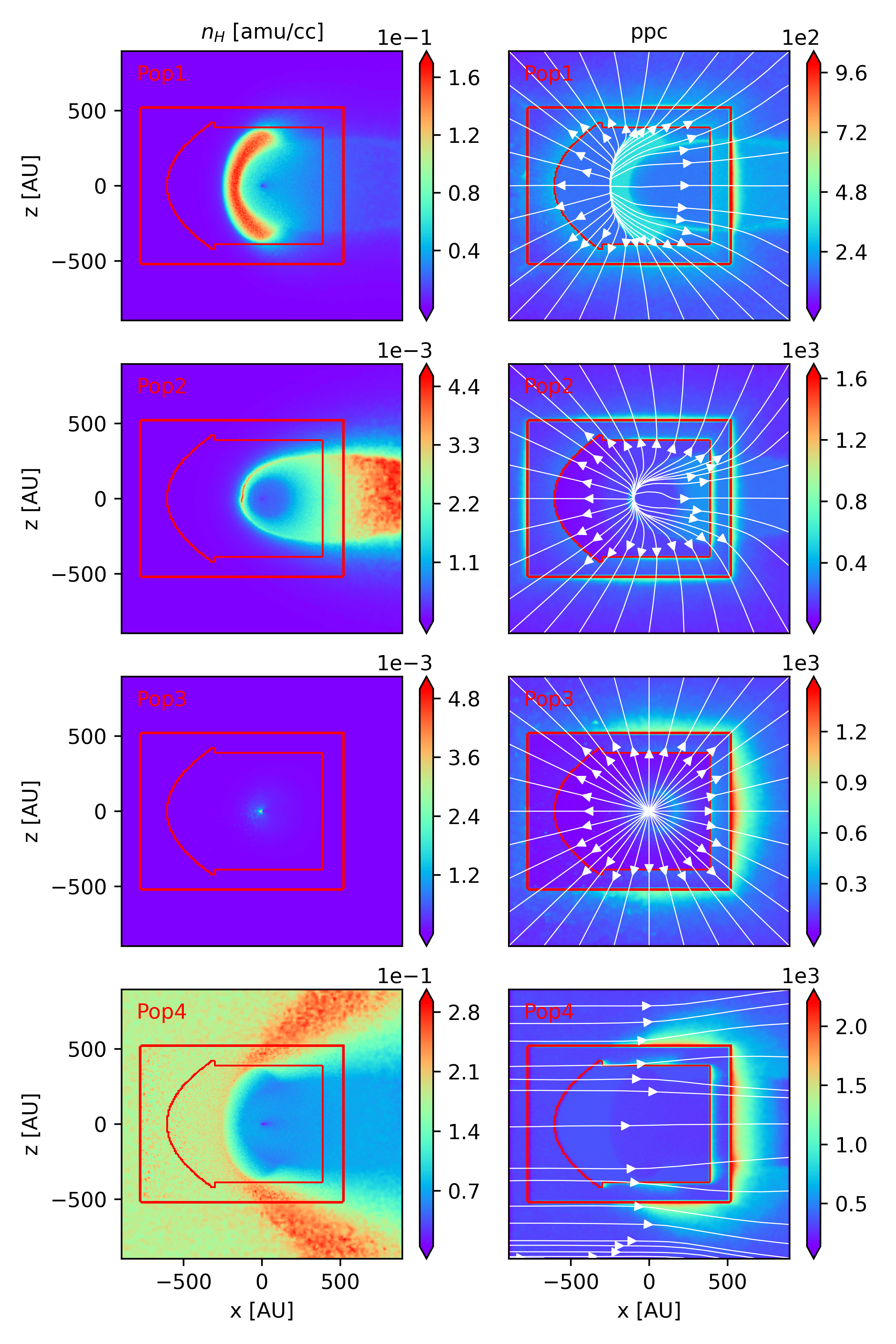}
    \caption{The number density (left column) and ppc number (right column) of the four populations in the $y=0$ plane. The red lines indicate the boundaries of AMR resolution change. The white lines are the streamlines of each population.}
    \label{fig:pops}
\end{figure}

\begin{figure}
    \centering
    \includegraphics[width=0.9\textwidth]{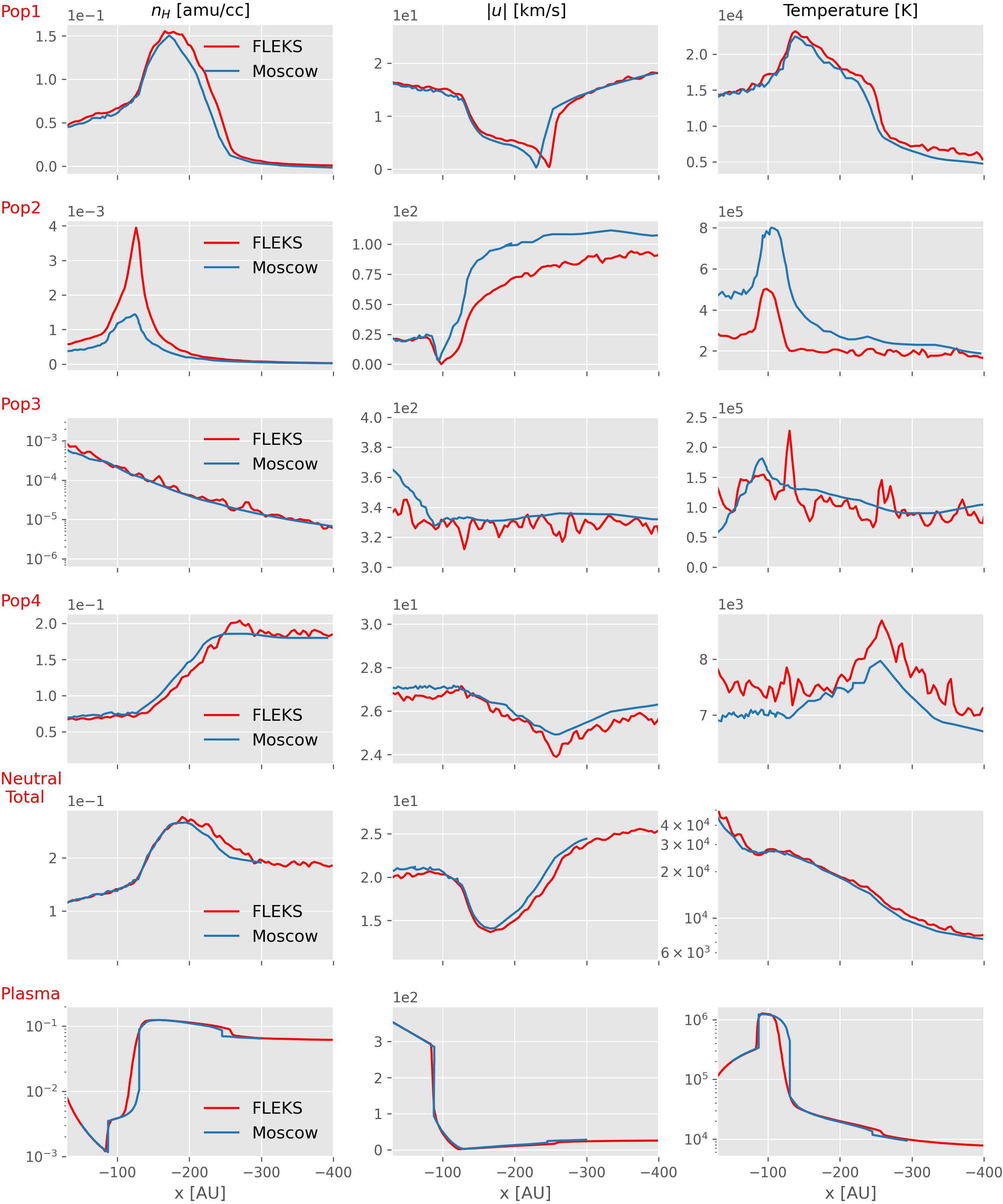}
    \caption{From left to right, the density, flow speed, and temperature of different species along the nose of the heliosphere, which is indicated by the black line in Figure~\ref{fig:amr}, are shown. The top four rows depict the properties of the four neutral populations, followed by the total neutral properties (row 5) and the plasma properties (row 6). The red lines represent our kinetic-MHD simulation results, and the blue lines are the results from \citet{alexashov_kinetic_2005}. For better visualization, the following quantities are shown in logarithmic scale: pop3 density, plasma density, total neutral temperature, and plasma temperature.}
    \label{fig:lines}
\end{figure}

To showcase the applicability of SHIELD-2 to more realistic scenarios, we simulate the same heliosphere case that has been presented in \citet{michael_solar_2022} and \citet{alexashov_kinetic_2005}, and show the comparison in this section. 

The incoming ISM flow parameters are: $n_{ISM} = 0.06\,$amu/cc, $T_{ISM} = 6519\,$K, and $\mathbf{u}_{ISM} = (26.4, 0, 0)\,$km/s. The supersonic solar wind parameters are: $n_{sw} = 0.00874\,$amu/cc, $T_{sw} = 1.126\times 10^5\,$K, and the velocity is radial with  $|\mathbf{u}_{sw}| = 354.75\,$km/s. 
The OH simulation domain is a box that spans from $(-1000, -1000, -1000)\,$AU to $(1000, 1000, 1000)\,$AU. The maximum grid resolution reaches $\Delta x = 1\,$AU near the Sun, and it is at least $\Delta x = 4\,$AU inside the box region from $(-400, -400, -400)\,$AU to $(400, 400, 400)\,$AU. As shown in Figure~\ref{fig:workflow}, we run the multi-fluid OH model first to obtain a steady-state solution for both the plasma and the neutrals, and then we start the kinetic-MHD simulation from the steady-state solution. 
The PT component covers the box region from $(-896, -896, -896)\,$AU to $(896, 896, 896)\,$AU with a based grid resolution of $\Delta x = 16\,$AU, and it is refined twice to $\Delta x = 4\,$AU in the inner region with a refinement ratio of 2, as shown in Figure~\ref{fig:amr}. The region with $\Delta x = 4\,$AU is defined by a combination of a box and a paraboloid. The PT time step is $\Delta t = 0.025\,$year, and OH and PT are coupled every 0.1 years. For a neutral with a speed of $400\,$km/s, which is about the fastest neutral speed in the simulation, it moves about 2\,AU, i.e., about half of a cell, within a single time step, so the time step is small enough to resolve the neutral motion. 
From the multi-fluid steady-state solution, we run the coupled simulation for another 200 years, which is 8000 steps for PT, and present the results in this section. 
In this simulation, the inner boundary for the OH component is set at $r=30\,$AU. There is no inner boundary for the PT component and the following plasma properties are used for simulating the charge exchange inside the sphere of $r<30\,$AU:
\begin{eqnarray}        
        n(r) &=& n_{sw}\cdot\left(\frac{30\,\mathrm{AU}}{r}\right)^2\\
        T(r) &=& T_{sw} \\
        \mathbf{u}(r) &=& |\mathbf{u}_{sw}|\mathbf{e}_r.   
\end{eqnarray}
Initially, $N_{init}=125$ macro-particles per population per cell are launched based on the steady-state multi-fluid simulation results. Since the mass density of a population can be very low in some areas, we set a minimum density threshold of $n_{H, vacuum} = 1 \times 10^{-4} \text{amu/cc}$. If the density of a population in a cell falls below this threshold, we do not launch any macro-particles for that population in that cell. This threshold is chosen to be three orders of magnitude lower than the typical total neutral density, ensuring a negligible impact on the solution. 

At most $N_{source} = 8$ new source macro-particles are generated for each cell at every step. The meaning of $N_{source}$ is described in section~\ref{sec:interaction}. If the ppc number exceeds the merging threshold $N_{high}$, we merge 10 macro-particles into 8. If the ppc number falls below the splitting threshold $N_{low}$ or some macro-particles have excessively high masses (section~\ref{sec:splitting}), we apply the particle splitting algorithm. Since we are particularly interested in the region with a high grid resolution, we aim to maintain a higher ppc number compared to the rest of the domain. Therefore, we use level-dependent thresholds: $N_{high} = 2N_{init}\cdot 1.5^L$ and $N_{low}=0.8N_{init} \cdot 1.5^L$, where $L$ is the refinement level (with $L = 0$ for the base grid).

The densities and ppc numbers for each population are shown in Figure~\ref{fig:pops}. The transition near the resolution change boundary is smooth, and there are no noticeable disruptions in the density distributions. 
When neutral particles flow from the high-resolution region to the coarse grid, the ppc number can significantly increase to about 2000 on the coarse side, as shown in the right column of Figure~\ref{fig:pops}. 
These macro-particles are then gradually merged as they move away from the refined region. Initially, there are about 75 macro-particles per cell per population on average, and this number increases to 198 at the end of the simulation. Without particle merging, this value would reach several thousand after 8000 steps. We note the initial average ppc number is 75 instead of $N_{init}=125$ because of the aforementioned vacuum threshold.

Figure~\ref{fig:lines} compares the simulation results for each population (first four rows), the total neutral (fifth row), and the plasma (last row) properties with those presented in \citet{alexashov_kinetic_2005} along the heliosphere nose (black line in Figure~\ref{fig:amr}). Although \citet{alexashov_kinetic_2005} employed a 2D axisymmetric simulation, while ours is 3D, both simulations are physically equivalent since there is no magnetic field in the simulations and the initial conditions and boundary conditions are axisymmetric. However, there are a few differences in the numerical implementations: 
\begin{itemize}
    \item Our inner boundary of the OH component is at $r=30\,$AU, whereas it is at $r=1\,$AU in \citet{alexashov_kinetic_2005}.
    \item The grid resolution and the numerical schemes differ between the simulations.
\end{itemize}
Therefore, we anticipate some discrepancies but not substantial deviations. The discrepancy in Pop2, where our simulation shows a higher neutral density, is likely caused by variations in population region definitions, as discussed in \citet{michael_solar_2022}. 
Pop3 neutrals, generated in the supersonic solar wind region and moving radially away from the Sun, exhibit a density reduction of about two orders of magnitude within 300 AU. This radial flow also leads to a decrease in ppc number and an increase in statistical noise. Despite the systematic difference in Pop2 and the statistical noise in Pop3, the plasma and total neutral properties agree with the \citet{alexashov_kinetic_2005} results very well. 
It is important to note that all the lines of FLEKS in Figure~\ref{fig:lines} are directly obtained from the final simulation results without any temporal or spatial averaging.

The 200-year simulation completes in approximately 50 hours using 2000 Intel Skylake CPUs. The kinetic PT component consumes roughly $90\%$ of the computational time, with the remaining $10\%$ attributed to the fluid OH component. With the capabilities of modern supercomputers, we can further increase the grid resolution for a production simulation if necessary.

\section{Summary}
\label{sec:summary}

In this work, we present our novel kinetic-MHD model SHIELD-2 for simulating neutral-ion interactions within the heliosphere.
The developement of this model is a critical advancement within the SHIELD DRIVE Science Center \citep{opher2023solar}. SHIELD-2 incorporates critical features such as Adaptive Mesh Refinement (AMR) and particle splitting/merging, enabling efficient 3D time-dependent simulations. Validation tests confirm the proper functioning of the AMR grid and particle resampling algorithms. Furthermore, the 3D simulation of the outer heliosphere shows excellent agreement with the results of \citet{alexashov_kinetic_2005}, demonstrating that SHIELD-2 accurately represents the essential charge exchange physics in the heliosphere. The efficiency of the 3D simulation demonstrates that it is feasible to perform time-dependent simulations with SHIELD-2.

In our current implementation, the neutral kinetic model is coupled to a single-fluid MHD model. Since the thermal ions and the pickup ions can have very different velocities and temperatures, a two-ion-fluid model, which includes both the thermal ions and the pickup ions\citep{opher2020small}, is crucial for accurate modeling of the heliosphere. We will extend our kinetic-MHD model to support pickup ions in the future.

\begin{acknowledgments}
This work is supported by NASA grant 18-DRIVE18\_2-0029, Our Heliospheric Shield, 80NSSC22M0164. Computational resources supporting this work were provided by the NASA High-End Computing (HEC) Program through the NASA Advanced Supercomputing (NAS) Division at Ames Research Center.
\end{acknowledgments}




\bibliography{csem,oh}{}
\bibliographystyle{aasjournal}



\end{document}